\documentclass{PoS}

\usepackage{amsmath}
\usepackage{amssymb}
\usepackage{cleveref}

\title{Lattice study of ChPT beyond QCD}

\ShortTitle{Lattice study of ChPT beyond QCD}

\author{ \speaker{Ethan~T.~Neil}${}^a$, Adam Avakian${}^b$, Ron Babich${}^b$, Richard~C.~Brower${}^b$, Michael Cheng${}^c$, Michael~A.~Clark${}^{fg}$, Saul~D.~Cohen${}^b$, George~T.~Fleming${}^a$, Joseph Kiskis${}^d$, James~C.~Osborn${}^e$, Claudio Rebbi${}^b$, David Schaich${}^b$, Pavlos Vranas${}^c$ \\
${}^a$ Department of Physics, Sloane Laboratory, Yale University,
             New Haven, CT 06520 \\
${}^b$ Department of Physics, Boston University,
	Boston, MA 02215 \\
${}^c$ Physical Sciences Directorate, Lawrence Livermore National Laboratory,
	Livermore, CA 94550 \\
${}^d$ Department of Physics, University of California,
	Davis, CA 95616 \\
${}^e$ Argonne Leadership Computing Facility,
	Argonne, IL 60439 \\
${}^f$  Harvard-Smithsonian Center for Astrophysics, Cambridge, MA 02138 \\
${}^g$ Initiative in Innovative Computing, Harvard University School of Engineering and Applied Sciences, Cambridge, MA 02138 \\
E-mail: \email{ethan.neil@yale.edu}}

\abstract{We describe initial results by the Lattice Strong Dynamics (LSD) collaboration of a study into the variation of chiral properties of $SU(3)$ Yang-Mills gauge theory as the number of massless flavors changes from $N_f = 2$ to $N_f = 6$, with a focus on the use of chiral perturbation theory.}

\FullConference{6th International Workshop on Chiral Dynamics\\
		 July 6-10 2009\\
		 Bern, Switzerland}

\begin{document}

\newcommand{\beq}{\begin{equation}}
\newcommand{\eeq}{\end{equation}}
\newcommand{\beqs}{\begin{eqnarray}}
\newcommand{\eeqs}{\end{eqnarray}}
\newcommand{\lsim}{\mathrel{\raisebox{-.6ex}{$\stackrel{\textstyle<}{\sim}$}}}
\newcommand{\gsim}{\mathrel{\raisebox{-.6ex}{$\stackrel{\textstyle>}{\sim}$}}}
\newcommand{\pbp}[0]{\ensuremath{\langle \overline{\psi} \psi \rangle}}
\newcommand{\chidof}{\ensuremath{\mbox{$\chi^2/\text{d.o.f.}$}}}

\bibliographystyle{abbrv}

\section{Motivation}

Theories different from QCD have seen a considerable amount of recent attention from the lattice gauge theory community.  In particular, the properties of $SU(N_c)$ Yang-Mills gauge theories with $N_f$ light fermions have been the subject of numerous recent lattice studies \cite{Appelquist:2007hu,Deuzeman:2008sc,Appelquist:2009ty,Deuzeman:2009mh,Fodor:2009wk} in the context of searching for novel non-perturbative physics which could play a role in building strongly-coupled models of physics beyond the standard model, e.g. technicolor.  We will consider here only the case $N_c = 3$, with $N_f$ fermions in the fundamental representation of the gauge group.

The infrared dynamics within this class of theories is known to undergo a transition as $N_f$ is varied within the range $0 < N_f < N_f^{af}$ for which asymptotic freedom is preserved.  It is well known that for $N_f$ just below $N_f^{af}$, the $\beta$-function of the theory shows a weak infrared (IR) fixed point \cite{Caswell:1974gg, Banks:1981nn}, leading to conformal behavior in the IR.  This is markedly different from the confining IR behavior of a QCD-like theory (small $N_f$); we expect a phase transition at some point as $N_f$ is varied between the two extremes.  Thus far, lattice studies have supported the picture that there is a ``conformal window" $N_f^c < N_f < N_f^{af}$, with the fixed-point coupling strength becoming strong as $N_f$ approaches the critical transition value $N_f^c$.

Lattice simulations are necessarily performed at finite fermion mass $m_f$, and must then be extrapolated to the physical point or the chiral limit.  Thus, chiral perturbation theory ($\chi$PT) is an invaluable tool in connecting our simulation results at various $N_f$ with continuum physics.  Although most interest thus far has been on $SU(2)$ and $SU(3)$ $\chi$PT for application to QCD, generalization of $\chi$PT to the breaking of an $SU(N_f) \times SU(N_f)$ chiral symmetry is quite straightforward, with the $N_f$ counting factors well-known up to next-to-leading order (NLO).  However, the values of the low-energy constants appearing in the chiral Lagrangian are not identical between the $SU(2)$ and $SU(N_f)$.

Of particular interest is the magnitude of the fermion condensate $\pbp$ relative to the Goldstone decay constant $F$, encapsulated in the ratio $\pbp / F^3$.  In extended technicolor models, there is significant tension between the requirements of evading experimental bounds on flavor-changing neutral currents (requiring a large ultraviolet cutoff $\Lambda_{ETC}$), and generating  standard-model quark masses which go as $\pbp_{TC} / \Lambda_{ETC}^2$, where $\pbp_{TC}$ is the technifermion condensate cut off at the scale $\Lambda_{ETC}$.  A phenomenon known as ``walking" can significantly enhance the ratio $\pbp / F^3$ relative to QCD \cite{Holdom:1981rm,Yamawaki:1985zg,Appelquist:1986an}, allowing generation of the correct masses without violating experimental bounds (see e.g. \cite{Lane:2000pa} for a brief review.)  Continuum studies based on Feynman graphs  have suggested just such a significant enhancement for theories with many flavors $N_f$, near (but just below) the transition value $N_f^c$ \cite{Lane:1991qh, Appelquist:1997fp}.

Determining the evolution of $\pbp/F^3$ is a primary goal of the Lattice Strong Dynamics (LSD) collaboration. Initial results were described in Ref. \cite{Appelquist:LSD}, comparing the well-known $N_f = 2$ theory to the case of $N_f = 6$, concluding that there is significant enhancement at $N_f = 6$. In this supplement to Ref. \cite{Appelquist:LSD}, to appear in the Proceedings of 6th International Workshop on Chiral Dynamics, we describe more completely the use of $\chi$PT and the extrapolation to $m = 0$, with particular focus on the difficulties that arise at $N_f = 6$.

In section 2, we review the $\chi$PT formulas for general $N_f$. In section 3, we detail our simulation methods and parameters. In section 4, we perform a variety of chiral fits to our data at both $N_f = 2$ and $N_f = 6$, with comparison between the two cases. In section 5, we make some concluding remarks.

\section{$\chi$PT at general $N_f$}

The application of $\chi$PT to theories with a general number of light fermion flavors $N_f$ (all with mass $m$) is straightforward.  Aside from the unknown $N_f$ dependence of the low-energy constants themselves, $N_f$ appears as a loop counting factor in NLO formulas and beyond.  The general formulas at NLO for $M_m^2$, $F_m$ and $\pbp_m$ are \cite{Gasser:1986vb}:
\begin{align}
\frac{M_m^2}{2m} &= B \left\{1 + \frac{2mB}{(4\pi F)^2} \left[2 \alpha_8 - \alpha_5 + N_f(2\alpha_6 - \alpha_4) + \frac{1}{N_f} \log \left(\frac{2mB}{(4\pi F)^2}\right) \right] \right\} \label{eq:mNLO} \\
F_m &= F \left\{1 + \frac{2mB}{(4\pi F)^2} \left[\frac{1}{2} (\alpha_5 + N_f \alpha_4) - \frac{N_f}{2} \log \left(\frac{2mB}{(4\pi F)^2} \right) \right] \right\} \label{eq:fNLO} \\
\pbp_m &= F^2 B \left\{ 1 + \frac{2mB}{(4\pi F)^2} \left[ \frac{1}{2} (2\alpha_8 + \eta_2) + 2N_f \alpha_6 - \frac{N_f^2 - 1}{N_f} \log \left(\frac{2mB}{(4\pi F)^2} \right) \right] \right\} \label{eq:cNLO}
\end{align}
where $\alpha_i \equiv 8 (4\pi)^2 L_i$ and $\eta_i \equiv 8 (4\pi)^2 H_i$ are just the conventional low- and high-energy constants of $\chi$PT, rescaled to values of $O(1)$.  Both the analytic terms and the chiral logarithms scale with the number of fermions $N_f$; except for the log correction to $M_m^2$, the size of the NLO terms increases relative to the leading order as $N_f$ increases.  The lone high-energy constant $\eta_2$ which appears in $\pbp_m$ includes a ``contact term", a quadratic divergence in the ultraviolet cutoff (here $1/a$, where $a$ is the lattice spacing).  We therefore expect the analytic term linear in $m$ to dominate the chiral expansion of $\pbp_m$.

Without fitting to additional observables or resorting to partial quenching, we cannot distinguish the individual low- and high-energy constants in \cref{eq:mNLO,eq:fNLO,eq:cNLO} above.  Therefore in our analysis, we adopt a simpler notation by combining many of the coefficients:
\begin{align}
\frac{M_m^2}{2m} &= 8\pi^2 F^2 z \left\{1 + zm \left[\alpha_{M} + \frac{1}{N_f} \log \left(zm\right) \right] \right\} \label{eq:mzNLO} \\
F_m &= F \left\{1 +zm \left[ \alpha_{F} - \frac{N_f}{2} \log \left( zm \right) \right] \right\} \label{eq:fzNLO} \\
\pbp_m &= 8 \pi^2 F^4 z \left\{ 1 + zm \left[ \alpha_{C} - \frac{N_f^2 - 1}{N_f} \log \left( zm \right) \right] \right\}, \label{eq:czNLO}
\end{align}
where we have also defined the parameter
\beq
\label{eq:zdef}
z \equiv \frac{2B}{(4\pi F)^2}.
\eeq
The quantity $zm$ is the expansion parameter of $\chi$PT, so that $1/z$ gives a rough estimate of the mass scale at which we expect perturbation theory to break down.

At next-to-next-to-leading order (NNLO), six additional coefficients appear, corresponding to the $m^2$ and $m^2 \log m$ terms in the chiral expansion.  The coefficients of the leading non-analytic terms $m^2 (\log m)^2$, like the $m \log m$ terms above, are completely determined by $N_f$, and have recently been computed for general $N_f$ \cite{Bijnens:2009qm}.  Extending the notation of \cref{eq:mNLO,eq:fNLO,eq:cNLO}, we write
\begin{align}
\frac{M_m^2}{2m} &= 8\pi^2 F^2 z \left\{1 + zm \left[\alpha_{M} + \frac{1}{N_f} \log \left(zm\right) \right] \right.\nonumber \\
&+\left. (zm)^2 \left[ \alpha_{M20} + \alpha_{M21} \log \left(zm\right) + \left(\frac{3}{8} N_f^2 + \frac{9}{2N_f^2} - \frac{1}{2} \right) \left(\log (zm)\right)^2 \right] \right\}, \label{eq:mzNNLO} \\
F_m &= F \left\{1 +zm \left[ \alpha_{F} - \frac{N_f}{2} \log \left( zm \right) \right] \right. \nonumber \\
&+\left. (zm)^2 \left[ \alpha_{F20} + \alpha_{F21} \log \left(zm\right) + \left( -\frac{3}{16} N_f^2 -\frac{1}{2} \right) \left(\log (zm)\right)^2 \right] \right\}, \label{eq:fzNNLO} \\
\pbp_m &= 8 \pi^2 F^4 z \left\{ 1 + zm \left[ \alpha_{C} - \frac{N_f^2 - 1}{N_f} \log \left( zm \right) \right] \right. \nonumber \\
&+\left. (zm)^2 \left[ \alpha_{C20} + \alpha_{C21} \log \left(zm\right) + \frac{3}{2} \left( \frac{1}{N_f^2} - 1 \right) \left(\log (zm)\right)^2 \right] \right\}. \label{eq:czNNLO}
\end{align}
Because of the dominance of the linear term in the expansion of $\pbp_m$, when fitting our data to \cref{eq:mzNNLO,eq:fzNNLO,eq:czNNLO} we will fix the unimportant NNLO analytic terms $\alpha_{C20} = \alpha_{C21} = 0$, in order to obtain a fit with more degrees of freedom.

\section{Simulation details}

The study of unexplored aspects of chiral dynamics is a difficult proposition on the lattice, as lattice fermion discretization typically breaks chiral symmetry explicitly.  For this study, we employ domain wall fermions \cite{Kaplan:1992bt, Shamir:1993zy}, which have good flavor symmetry properties and break chiral symmetry only by an exponentially small factor.  In addition to the domain wall fermion action, we use the Iwasaki improved gauge action \cite{Iwasaki:1985we}.

Gauge configurations are generated using the CPS application package, part of the USQCD software library.  Evolution in configuration space is performed via the hybrid Monte Carlo (HMC) method, optimized with a three-level symplectic integrator, a single level of Hasenbusch preconditioning, and chronological inversion.

The lattice volume is set to $32^3 \times 64$.  For the length of the fifth dimension we take $L_s = 16$, and the domain-wall height is set to $m_0 = 1.8$.  Input fermion masses are varied from $m_f = 0.005$ to $0.03$.  Raw data are blocked over sets of 50 trajectories before analysis, in order to reduce the effects of autocorrelations; our runs are not long enough to perform a complete analysis of autocorrelation times.

Since we do not extrapolate $L_s \rightarrow \infty$, there is some residual chiral symmetry violation, which can be encapsulated in a residual mass $m_{res}$.  The magnitude of the residual mass in the chiral limit is $m_{res} = 2.60 \times 10^{-5}$ in the $N_f  = 2$ theory, and $m_{res} = 8.23 \times 10^{-4}$ for $N_f = 6$.  To take the chiral limit, we extrapolate in the total fermion mass $m \equiv m_f + m_{res}$.  The variation of $m_{res}$ over the range of $m_f$ used is extremely small compared to $m$, and we neglect it here.

In order to obtain a correct sampling of the partition function in finite volume, evolution of global topological charge $Q$ is important.  Our lightest evolutions at $m_f = 0.005$ show signs of very slow evolution in topological charge, which can lead to significant systematic shifts, particularly in $\pbp_m$ and $F_m$ \cite{Leutwyler:1992yt}.  Due to this unknown systematic effect, we exclude the $m_f = 0.005$ results wherever possible below, and caution that any fits which do contain data at this mass point are questionable.  

\section{Chiral extrapolation}

We will consider an assortment of chiral fits, at both NLO and NNLO, varying both $N_f$ and the mass ranges included in the fit.  The set of fits to be considered is detailed in \Cref{table:types}.  Resulting best-fit parameters are shown in \Cref{table:params}.

\begin{table}[b]\begin{center}
\begin{tabular}{|c|c||c|c|c||c|}
\hline
Fit label&Order&$m_f$ range ($M_m^2$)&$m_f$ range ($F_m$)&$m_f$ range ($\pbp_m$)&$N_{dof}$\\
\hline
A&NLO&0.01-0.02&0.01-0.02&0.01-0.02&4\\
B&NLO&0.005-0.02&0.005-0.02&0.005-0.02&7\\
C&NLO&0.01-0.02&0.01&0.01&0\\
D&NNLO&0.005-0.02&0.005-0.02&0.005-0.02&3\\
E&NNLO&0.01-0.03&0.01-0.03&0.01-0.03&6\\
\hline
\end{tabular}\end{center}
\caption{\label{table:types} Types of chiral fit considered.  Both the order of the fit and mass ranges of data included for $M_m^2$, $F_m$ and $\pbp_m$ data are varied.  Fits A through D are performed for both $N_f = 2$ and $N_f = 6$; fit E can be performed only at $N_f = 6$, where data at $m_f = 0.025, 0.030$ are available. }
\end{table}

We begin at $N_f = 2$ with fit type A, which is simply an NLO fit to all of the available data (excluding $m_f = 0.005$, for the reasons given above.)  The $\chidof$ is 6.5 with 4 degrees of freedom; given that our error bars are most likely underestimated due to the relatively small number of trajectories gathered, this indicates a reasonably good fit to the data.  From \Cref{table:params} we note that $\alpha_M$ and $\alpha_F$ are $O(1)$ numbers, while $\alpha_C$ is very large so that the linear term dominates the $\pbp_m$ extrapolation, as expected from the presence of the ``contact term".  We expect the breakdown of chiral perturbation theory for $zm \approx 1$; since $1/z = 0.036$, the expansion should be good over the fit range used.

Despite our concerns about the $m_f = 0.005$ data, a natural question at this point is whether the NLO fit can be extended to include them.  We attempt fit type B in order to investigate.  The $\chidof$ rises significantly to 36.2, now with 7 degrees of freedom.  The inability to incorporate these lighter-mass points into the original chiral fit is most likely due to the presence of large systematic errors in the data which are not reflected in the error bars, as we anticipated.

\begin{figure}\begin{center}$\begin{array}{cc}
\includegraphics[width=75mm]{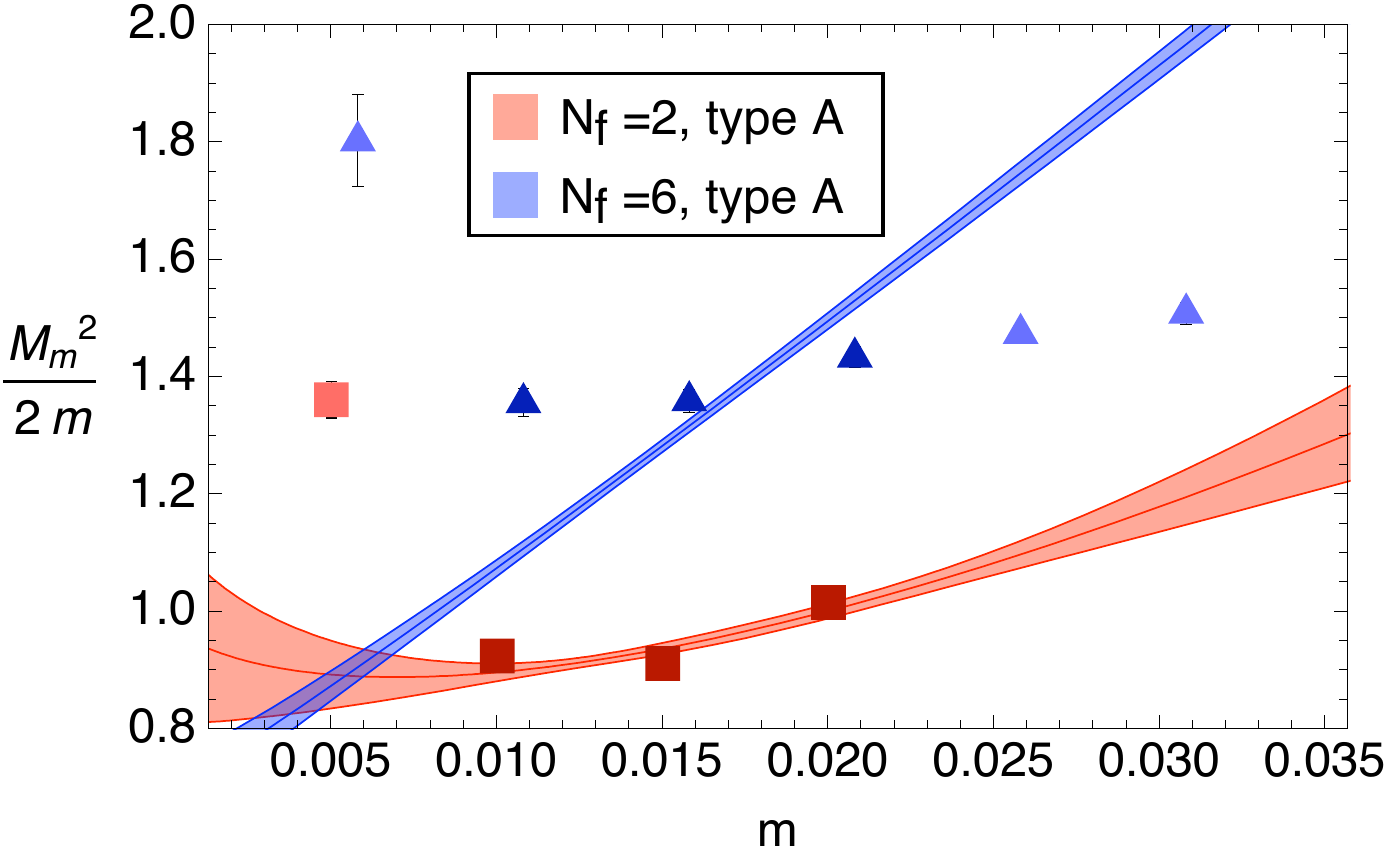}&
\includegraphics[width=75mm]{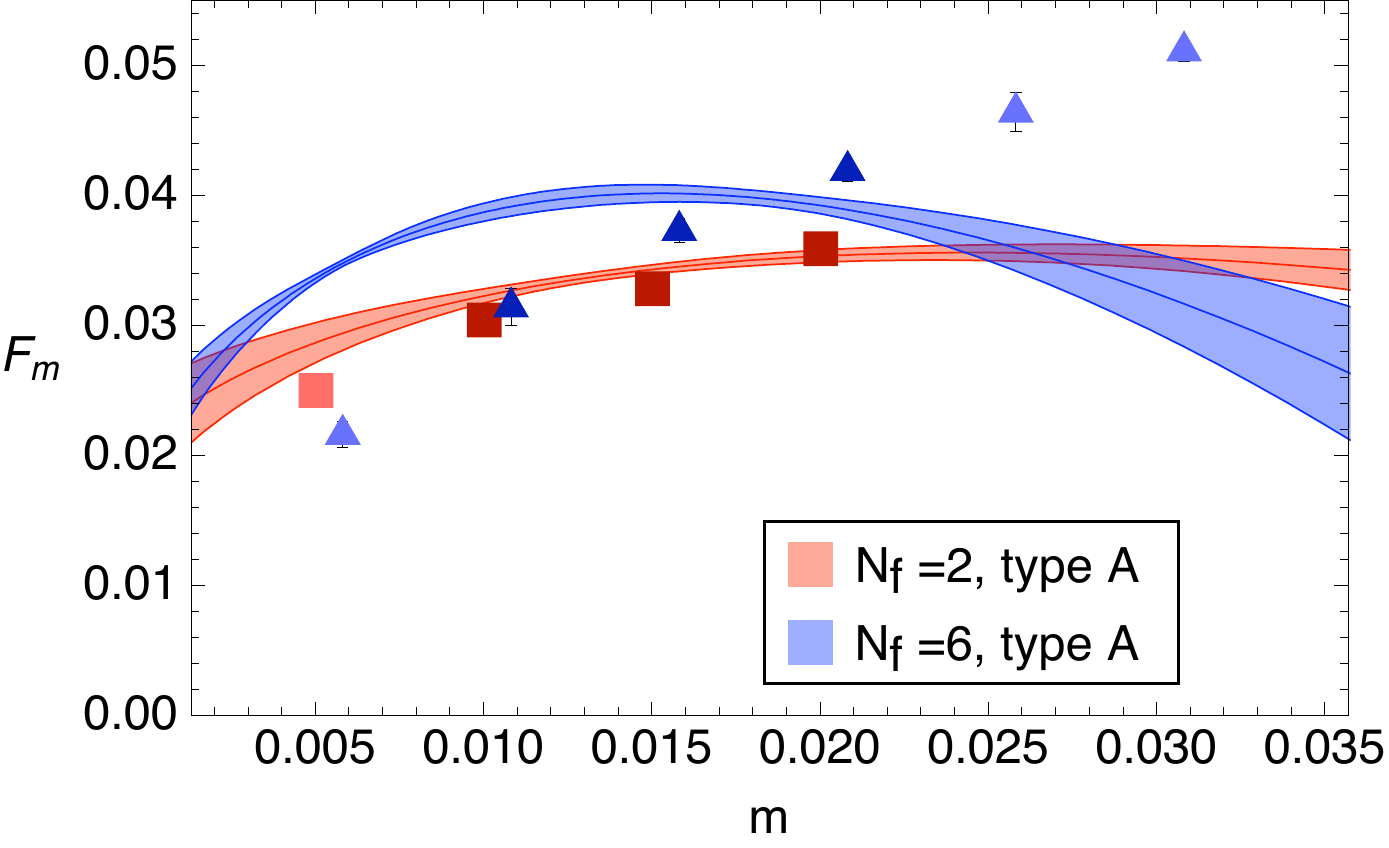}\end{array}$
\caption{\label{fig:type_A}
Chiral extrapolation of the quantities $M_m^2/2m$ and $F_m$, with $\chi$PT fits of type A.  Data points used in deriving the fits shown are dark red squares ($N_f = 2$) and dark blue triangles ($N_f = 6$); data excluded from the fit are lightly shaded.  Best-fit results with 1-$\sigma$ error bands are shown in corresponding colors.
}\end{center}
\end{figure}

Attempting the same basic NLO fit (type A) at $N_f = 6$ is the logical next step, but unfortunately, the data does not support such a fit.  As shown in \Cref{table:params}, the values of the best-fit parameters are fairly close to those from the $N_f = 2$ fit, but the $\chidof$ is now 50.5, indicating a very poor fit to the data.  Both fits of type A are plotted in \Cref{fig:type_A}, and the poor quality of the $N_f = 6$ fit is immediately obvious.  There is clear tension in the fit between the intercept values $F$ and $B$ expected from more naive extrapolation of the data, and the magnitude of the NLO terms governed by $z \propto B / F^2$.  Including the $m_f = 0.005$ data (type B) does not improve the situation, with roughly the same best-fit parameters and a somewhat larger $\chidof$.

\begin{figure}\begin{center}$\begin{array}{cc}
\includegraphics[width=75mm]{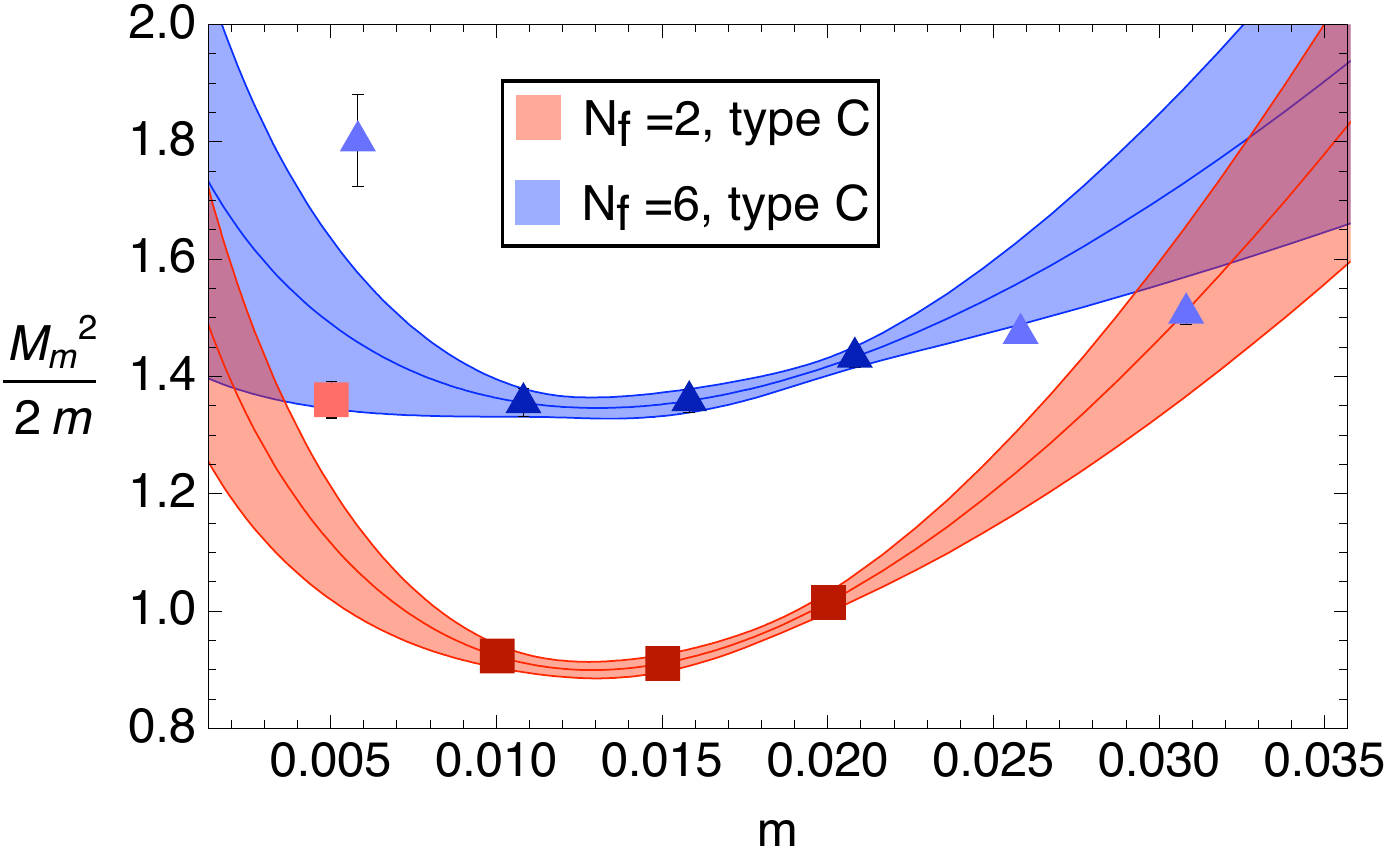}&
\includegraphics[width=75mm]{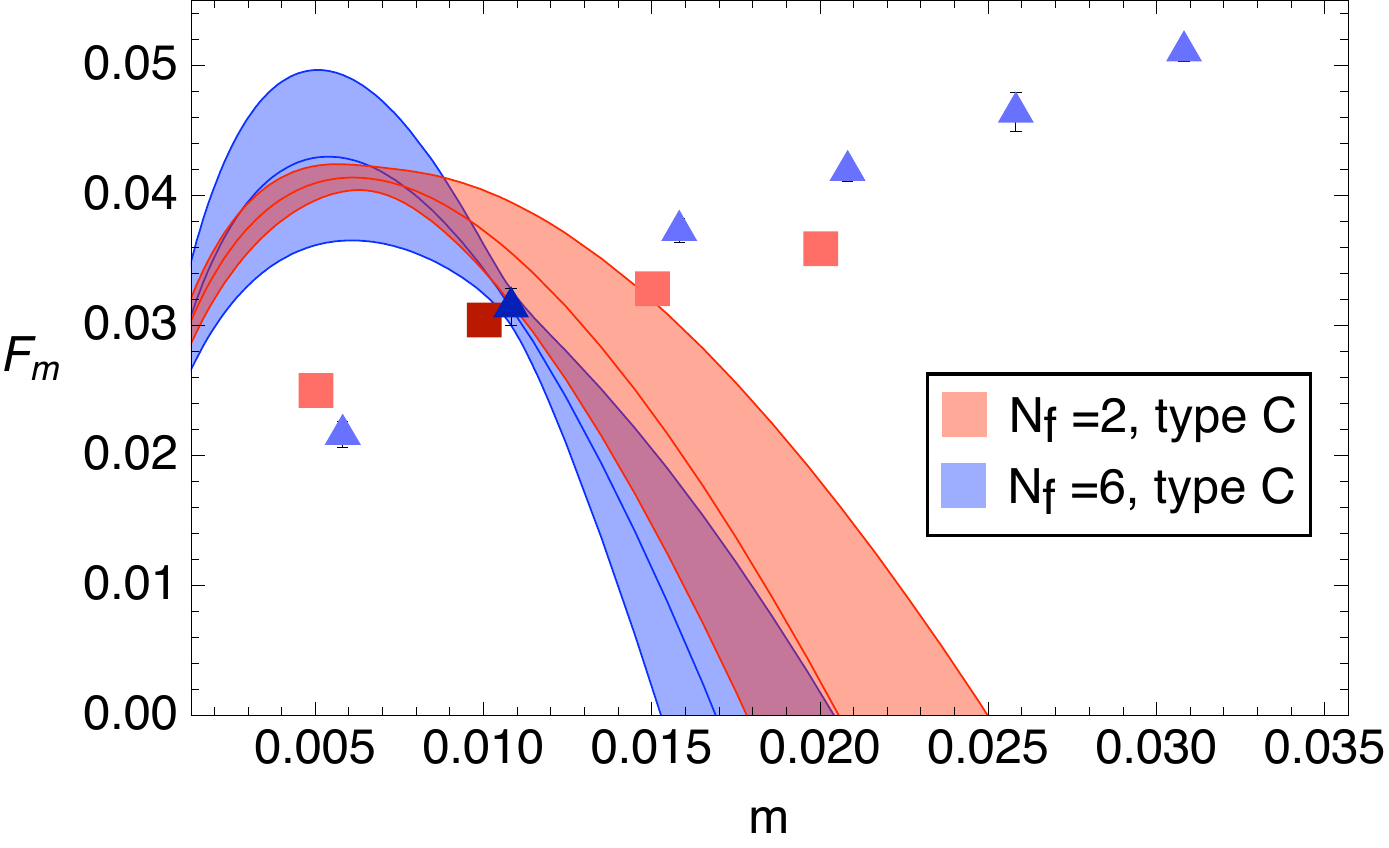}\end{array}$
\caption{\label{fig:type_CD}
Chiral extrapolation of the quantities $M_m^2/2m$ and $F_m$, with $\chi$PT fits of type C at both $N_f = 2$ (red) and 6 (blue).  Data symbols and colors are as in \Cref{fig:type_A}.
}\end{center}
\end{figure}

\begin{figure}\begin{center}$\begin{array}{cc}
\includegraphics[width=75mm]{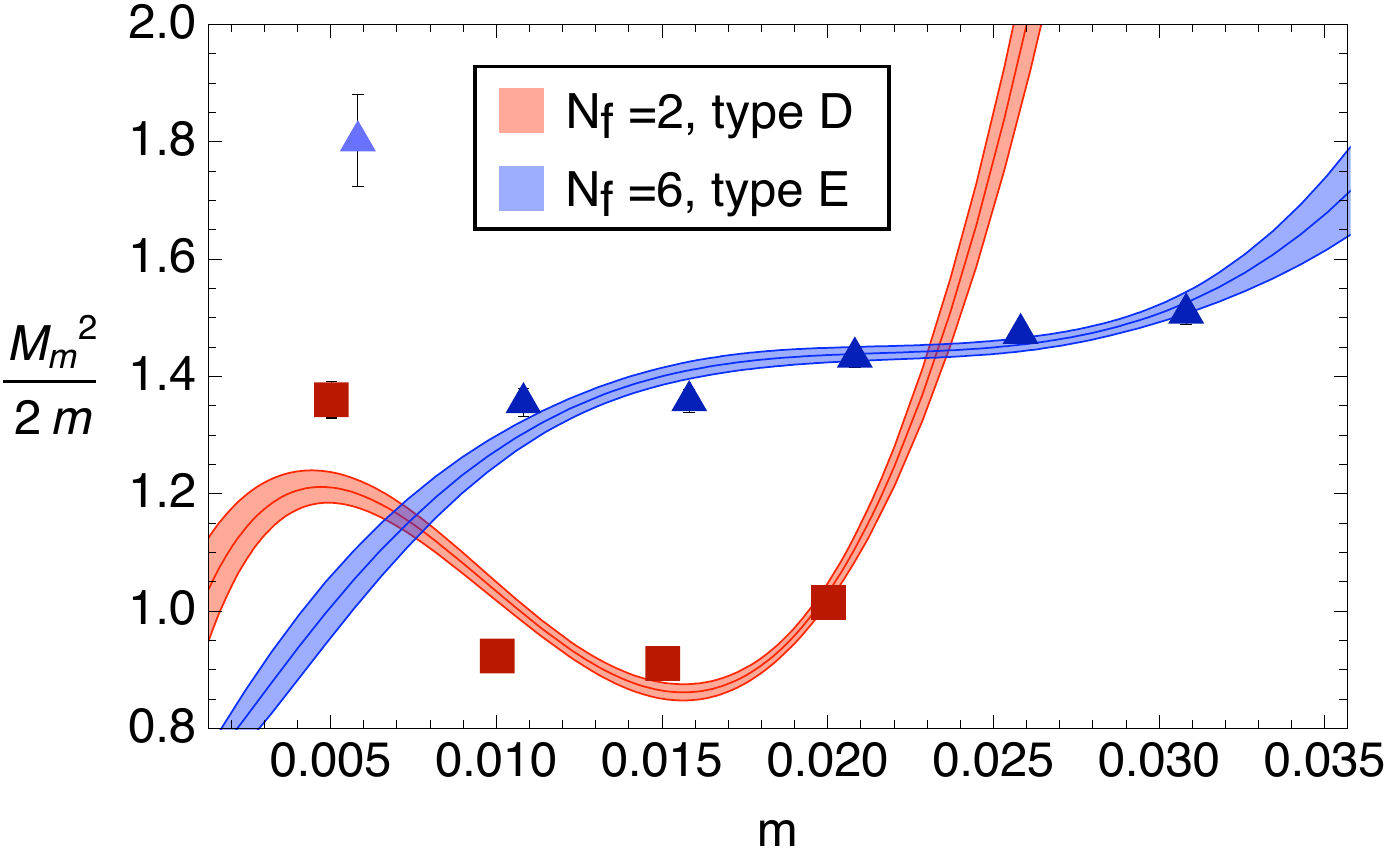}&
\includegraphics[width=75mm]{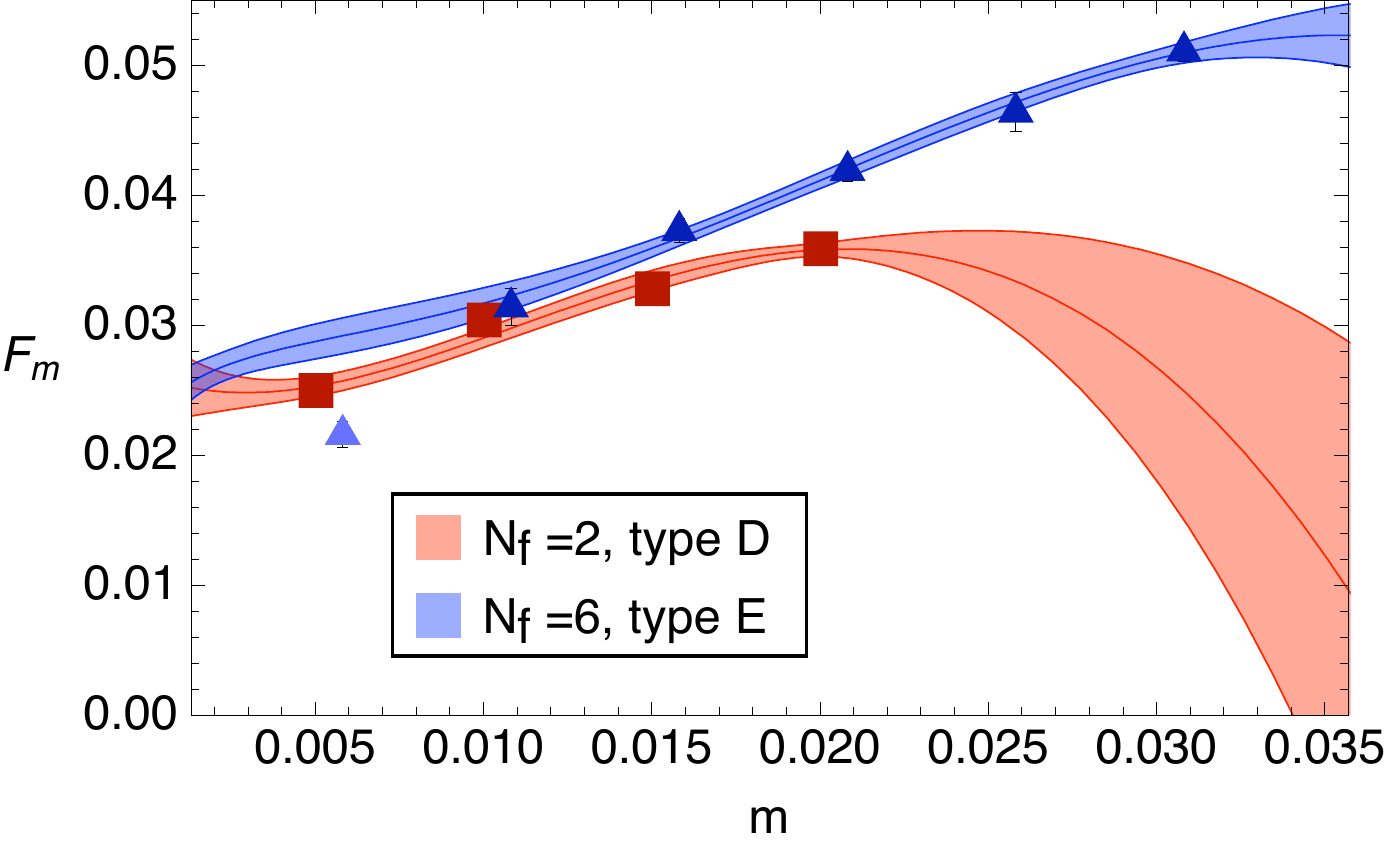}\end{array}$
\includegraphics[width=85mm]{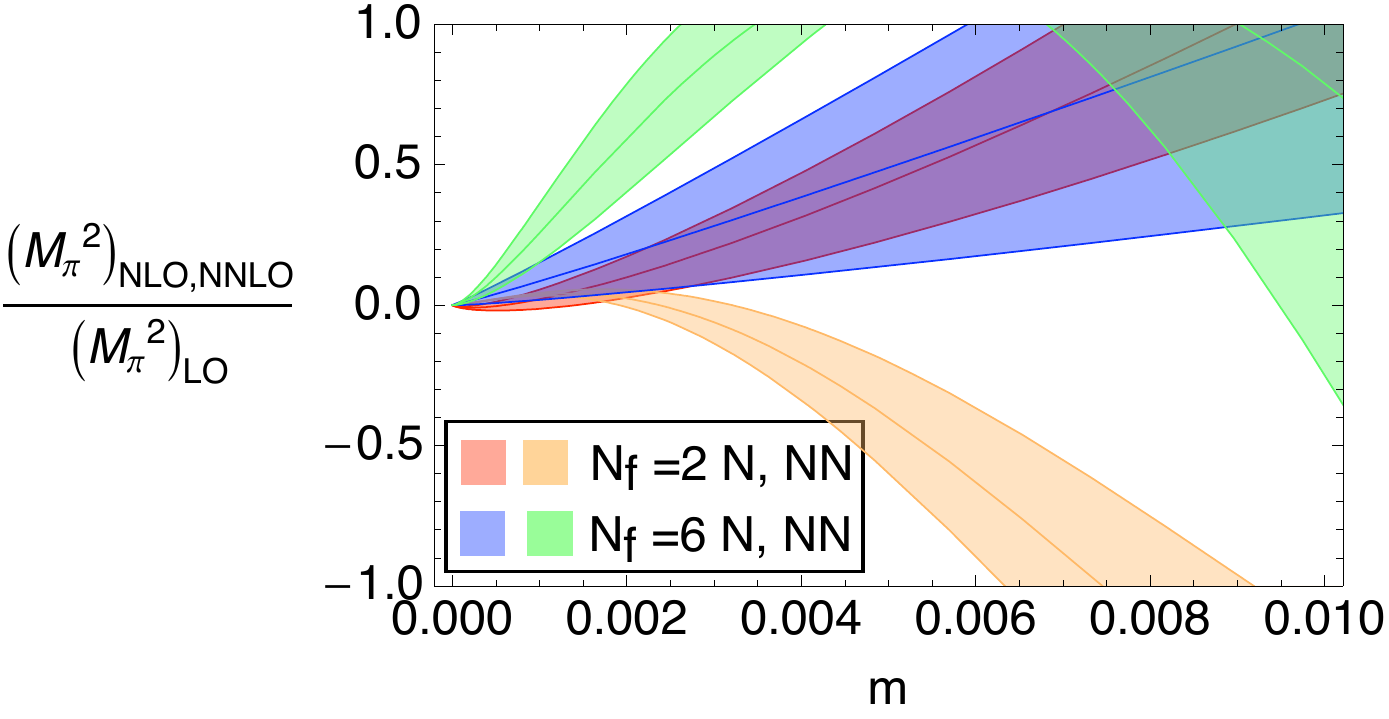}
\caption{\label{fig:type_EF}
NNLO chiral extrapolation of the quantities $M_m^2/2m$ and $F_m$, with fits of type D at $N_f = 2$ and type E at $N_f = 6$.  Data symbols and colors are as in \Cref{fig:type_A}.  The ratios of NLO (red: $N_f = 2$, blue: $N_f = 6$) and NNLO (orange: $N_f = 2$, green: $N_f = 6$) terms to LO in the $F_m$ extrapolation are also displayed, showing that all terms become quite large relative to LO near the bottom of the mass range used.
}\end{center}
\end{figure}

The failure of NLO $\chi$PT at $N_f = 6$, over the same range of $m_f$ which yielded a good fit at $N_f = 2$, is not entirely surprising in retrospect.  Most of the NLO terms in \cref{eq:mNLO,eq:fNLO,eq:cNLO} scale linearly with $N_f$, so that if we keep $z$ fixed, then as we go from $N_f = 2$ to 6, we expect to have to work in a mass range which is a factor of $3$ smaller to keep the size of the NLO terms fixed relative to LO.

An exception is the slope of the Goldstone mass squared, \cref{eq:mNLO}.  There, the chiral log is actually suppressed by $1/N_f$ (although there is still a linear increase with $N_f$ buried in the linear coefficient $\alpha_M$.)  Indeed, by inspection the variation of $M_m^2/2m$ over the mass range $0.01 \leq m_f \leq 0.02$ is quite small, both at $N_f = 2$ and $N_f = 6$.  We might therefore expect that a chiral fit to just the Goldstone mass data might be more successful.  This is fit type C, which includes the range of $M_m^2$ data $0.01 \leq m_f \leq 0.02$, along with the lightest data point $m_f = 0.01$ for both $F_m$ and $\pbp_m$ in order to fix $\alpha_F$ and $\alpha_C$.  We have zero degrees of freedom, so evaluation of goodness of fit is difficult.

Results for fits of type C are shown in \Cref{fig:type_CD}.  Both fits of type C lead to a large value of the parameter $z$, pointing to an early breakdown of the chiral expansion; in both cases, the NLO terms in $F_m$ are clearly large compared to the leading-order value at quite small $m$, with fine-tuned cancellation of the NLO terms occurring in order to match the single $F_m$ point contained in the fits.  Although we have no $\chidof$, the type C fit curves clearly do not match the $F_m$ data except at the point $m_f = 0.010$ included in the fit.

Finally, we attempt the NNLO fits of \cref{eq:mzNNLO,eq:fzNNLO,eq:czNNLO}.  As discussed we fix $\alpha_{C20} = \alpha_{C21} = 0$, so that there are 9 free parameters.  In order to obtain a constrained fit, we must include the $m_f = 0.005$ data at $N_f = 2$ (fit type D.)  At $N_f = 6$, we can use the range $0.01 \leq m_f \leq 0.03$ (fit type E), which we expect to yield a better result due to the systematic effects in the $m_f = 0.005$ data.  Results are plotted in \Cref{fig:type_EF}.  Fit type D, much like type B, includes the spurious $m_f = 0.005$ data, so the resulting large values of $\chidof$ are not surprising. Fit type E at $N_f = 6$ yields an apparently good fit to the data based on $\chidof$.  However, separating out the NLO and NNLO terms relative to the leading order shows that both terms are equal to or larger than the leading order even at the bottom of the mass range being fit to (\Cref{fig:type_EF}).  In addition, the analytic terms are generally large and strongly covariant.  The apparent goodness of fit is thus inconsistent with a convergent chiral expansion, despite the value of $\chidof$.

\begin{table}[b]\begin{center}
\begin{tabular}{|cc||c|c|c|c|c|c|}
\hline
$N_f$&label&$z$&$F$&$\alpha_M$&$\alpha_F$&$\alpha_C$&$\chidof$\\
\hline
2&A&28(16)&0.0209(4)&0.31(62)&0.64(47)&83(29)&6.50\\
6&A&25(11)&0.0188(36)&2.5(1.4)&0.1(1.1)&194(24)&50.5\\
2&B&44(9)&0.0184(13)&-0.10(11)&0.90(15)&58(6)&36.2\\
6&B&27(10)&0.0179(29)&2.4(1.2)&0.20(96)&204(21)&74.2\\
2&C&77(12)&0.0171(5)&-0.50(10)&0.74(11)&25(10)&$\infty$\\
6&C&138(63)&0.0133(14)&-0.26(10)&2.1(1.1)&28(10)&$\infty$\\
2&D&16(8)&0.0259(39)&20(18)&-6.6(5.9)&119(45)&30.0\\
6&D&17(13)&0.0214(66)&25(37)&-14(14)&237(73)&14.9\\
6&E&17.9(5.5)&0.0217(28)&1.8(7.6)&-2.2(4.0)&206(20)&6.08\\
\hline\hline
$N_f$&label&$\alpha_{M20}$&$\alpha_{M21}$&$\alpha_{F20}$&$\alpha_{F21}$&$\alpha_{C20}$&$\alpha_{C21}$\\
\hline
2&D&32(77)&81(104)&-5(17)&-24(29)&0(--)&0(--)\\
6&D&21(83)&92(167)&-7(41)&-51(74)&0(--)&0(--)\\
6&E&4.6(1.3)&14(18)&1.7(0.8)&-10(11)&0(--)&0(--)\\
\hline

\end{tabular}\end{center}
\caption{\label{table:params} Chirally extrapolated quantities and fit parameters, based on the assorted $\chi$PT fits considered.  All fits shown are joint fits between the three quantities $\pbp_m$, $F_m$ and $M_m^2$.  NNLO analytic coefficients $\alpha_{C20}, \alpha_{C21}$ are fixed to zero, due to the dominance of $\alpha_C$.}
\end{table}

\section{Conclusion}

Based on initial results from the LSD collaboration, we conclude that NLO $\chi$PT gives a self-consistent fit to our $N_f = 2$ data, but does not yield a satisfactory fit to the data at $N_f = 6$ for our current mass range $0.01 \leq m_f \leq 0.02$.  Alternative fit ranges and the inclusion of NNLO terms also fail to give an acceptable $N_f = 6$ fit.  As noted above, due to the linear scaling with $N_f$ of several terms we might expect that at lighter mass points (by a factor of 3) will be necessary in going from $N_f = 2$ to $N_f = 6$.  The inclusion of known finite-volume and finite-topology terms \cite{Leutwyler:1992yt} in the chiral fits should allow us to fit to lighter masses without systematic errors and without greatly increasing the computational cost.  Also, simulation at additional masses, particularly lighter ones, may allow for a better NNLO chiral fit which is consistent with a good chiral expansion.

Alternative approaches, including partially quenched analysis or the use of eigenvalue methods, may provide another way forward for the extraction of chiral quantities at large $N_f$, and will be explored by our collaboration in future studies.  Detailed analysis of the chiral extrapolation for the quantity $\pbp / F^3$ and of the physical implications of our results are carried out in \cite{Appelquist:LSD}.

We thank LLNL and the Multiprogrammatic and Institutional Computing program
for time on the BlueGene/L supercomputer. This work was supported partially by DOE grants DE-FG02-92ER-40704 (T.A., E.N.), DE-FG02-91ER40676 and DE-FC02-06ER41440, and NSF grants OCI-0749300, DGE-0221680, PHY-0427646 (A.A., R.B., R.C.B., M.A.C., S.D.C., C.R., D.S.), PHY-0835713 (M.A.C), and PHY-0801068 (G.F.). It was also supported by the DOE through ANL under contract DE-AC02-06CH11357 (J.O.), and the DOE Office of High Energy Physics through LLNL under contract DE-AC52-07NA27344 (M.C., P.V.).

\pagebreak

\end{document}